\documentclass[a4paper,useAMS,usenatbib]{mn2e}
\usepackage[totalwidth=480pt, totalheight=680pt]{geometry}
\usepackage{graphicx}

\def\gr{{$\gamma$-ray}}

\newcommand{\lsim}{{\lower.5ex\hbox{$\; \buildrel < \over \sim \;$}}}
\newcommand{\gsim}{{\lower.5ex\hbox{$\; \buildrel > \over \sim \;$}}}
\def\nup{$\nu_{\rm peak}^S$}

\newcommand{\fermi}{{\it Fermi}}
\newcommand{\fermibf}{{\it \textbf {Fermi}}}

\newcommand{\paperone}{{Paper I}}

\title[A simplified view of blazars: the \gr\ case]{A simplified view of blazars: the \gr\ case.
 }
 \author[P. Giommi et al.]{P. Giommi$^{1,3}$\thanks{E-mail:
paolo.giommi@asdc.asi.it}, P. Padovani$^{2,3}$, G. Polenta$^{1,4}$\\
$^{1}$ASI Science Data Center, c/o ESRIN, via G. Galilei, I-00044 Frascati, Italy \\
$^{2}$European Southern Observatory, Karl-Schwarzschild-Str. 2,
D-85748 Garching bei M\"unchen, Germany\\
$^{3}$Associated to INAF - Osservatorio Astronomico di Roma, via Frascati 33, I-00040
Monteporzio Catone, Italy\\
$^{4}$INAF - Osservatorio Astronomico di Roma, via Frascati 33, I-00040
Monteporzio Catone, Italy}

\begin{document}

\date{Accepted ... Received ...; in original form ...}

\pagerange{\pageref{firstpage}--\pageref{lastpage}} \pubyear{2012}

\maketitle

\label{firstpage}

\begin{abstract}
  We have recently proposed a new simplified scenario where blazars are classified as
  flat-spectrum radio quasars (FSRQs) or BL Lacs according to the prescriptions of unified schemes, and to a varying
  combination of Doppler boosted radiation from the jet, emission from the
  accretion disk, the broad line region, and light from the host galaxy.
  Here we extend our approach, previously applied to radio and X-ray
  surveys, to the \gr\ band and, through detailed Monte Carlo simulations,
  compare our predictions to \fermi-LAT survey data. Our simulations are in
  remarkable agreement with the overall observational results, including the
  percentages of BL Lacs and FSRQs, the fraction of redshift-less objects,
  and the redshift, synchrotron peak, and \gr\ spectral index distributions.
  The strength and large scatter of the oft-debated observed \gr\ -- radio flux density 
  correlation is also reproduced.
  In addition, we predict that almost 3/4 of \fermi-LAT BL Lacs, and basically all
  of those without redshift determination, are actually FSRQs with their
  emission lines swamped by the non-thermal continuum and as such should be
  considered. Finally, several of the currently unassociated high Galactic latitude
  \fermi\ sources are expected to be radio-faint blazars displaying a pure elliptical
  galaxy optical spectrum.

\end{abstract}


\section{Introduction}\label{intro}

Blazars are a small subclass of Active Galactic Nuclei (AGN) characterized
by distinctive and extreme observational properties, such as large
amplitude and rapid variability, superluminal motion, and strong emission
over the entire electromagnetic spectrum. These peculiar sources are known
since the discovery of QSOs \citep{schmidt63}, as 3C273, the first quasar
to be discovered, is in fact a well known blazar. A major distinction
between QSOs and blazars is that while the former type of AGN mostly emit
radiation through accretion onto a supermassive black hole, blazars also
host a jet that is oriented closely along the line of sight, within which
relativistic particles radiate losing their energy
in a magnetic field \citep{bla78,UP95}. This jet component is responsible for the non-thermal 
emission from radio to high-energy $\gamma$-rays and for the extreme properties listed above 
when the viewing angle is small. When this angle is larger than $15-20^{\circ}$ the
relativistic amplification and superluminal effects are much less important and jet emission 
across the electromagnetic spectrum, although still present, may not be dominating the flux at IR, optical and X-ray frequencies 
where other nuclear and non-nuclear components emerge. Under these conditions the source is usually called a radio galaxy.

Although blazars represent a small minority of AGN,
the interest in these sources is growing as they are being
found in increasingly large numbers in high Galactic latitude surveys
performed at microwave and \gr\ energies \citep{GiommiWMAP09,fermi1lac,
PlanckERCSC}. Blazars represent also the most abundant population of extragalactic sources 
at TeV energies\footnote{http://tevcat.uchicago.edu/}.

Due to the blend of accretion  disk, jet and host galaxy emission that is often
present in their optical spectrum, the classification of blazars based on
observational properties is a complex issue. This has resulted in some
confusion in the literature.

In a recent paper \citet[][hereafter Paper I]{paper1} examined the various
argumentations and the experimental evidence behind the classification of
blazars that have appeared in the literature, and proposed a new scenario,
which is based on optical/UV light dilution, minimal assumptions on the physical
properties of the non-thermal-jet emission, and unified schemes. These
posit that BL Lacs and flat-spectrum radio quasars (FSRQs) are mostly 
low-excitation (LERGs)/Fanaroff-Riley (FR) I and high-excitation (HERGs)/FR
II radio galaxies with their jets forming a small angle with respect to the
line of sight \citep[see][and references therein, for a discussion on LERGs
and HERGs]{but10}. We call this new approach the \textit {simplified blazar view}.
 By means of detailed Monte Carlo simulations, Paper I showed
that this scenario is consistent with the complex observational
properties of blazars as we know them from all the surveys carried out so
far in the radio and X-ray bands, solving at the same time a number of
long-standing issues. 
 
 \begin{table*}
 \caption{The  {\textit {simplified blazar view} scenario.  See  text for details.}}
 \begin{tabular}{lccl}
       & LERG & HERG & viewing angle \\
  \hline
 strong jet dilution (EW $< 5~$\AA; Ca H\&K $< 0.4$) &  BL Lac  & {\it BL Lac}$^{(1)}$  & $\theta < \theta_{\rm blazar}$  \\
 weak jet dilution & {\it radio galaxy}$^{(2)}$ & FSRQ &  $\theta < \theta_{\rm blazar}$ \\
 misdirected jet   &  radio galaxy &  radio galaxy &  $\theta > \theta_{\rm blazar}$ \\
 \hline
 \multicolumn{4}{l}{\footnotesize {\it Italics} denote ``masquerading" sources (see text), $^{(1)}$ misclassified FSRQ,  $^{(2)}$ misclassified BL Lac}
  \end{tabular}
 \label{tab:scenario}
 \end{table*}

 In this paper we extend the Monte Carlo simulations presented in
 \paperone\ to the GeV \gr\ band and compare our expectations with the AGN
 data in the \fermi-LAT 2-yr source catalogue \citep[][hereafter
 2LAC]{fermi2fgl,fermi2lac,shaw13} and the \fermi-LAT data of a sample of radio
 selected blazars \citep{GiommiPlanck,RadioPlanck}. We note that
 simulations in the \gr\ band are more complex than those in the X-ray band
 since the \gr\ flux of blazars is not due to simple synchrotron or 
 synchrotron-self-Compton emission, but is usually attributed to inverse
 Compton scattering of synchrotron and of external photons, or to the
 superposition of more than one component. Furthermore, the $\gamma$-ray 
 flux is a much broader-band measurement and is subject to the highest source 
 variability.

 Table \ref{tab:scenario} summarizes the {\textit {simplified blazar view}  proposed in
 \paperone. Sources with their jets forming small angles w.r.t. the line
 of sight \citep[$\theta < \theta_{\rm blazar} \sim 15 -
 20^{\circ}$;][]{UP95}, and therefore dominated by non-thermal emission,
 are characterized by low values of the equivalent width (EW) and/or the so-called
 Ca H\&K break, the latter being a stellar absorption feature found in the spectra of
 elliptical galaxies. However, only LERGs belonging to this class are real
 BL Lacs, that is have intrinsically weak emission lines. HERGs, which are
 supposed to have an accretion disk and therefore to display strong
 emission lines, {\it appear} to show weak lines only because these are
 swamped by non-thermal emission and are therefore ``masquerading" BL Lacs 
 ({\it  italics}  in the table) and are in fact misclassified FSRQs. Sources with somewhat weaker jet dilution still oriented
 at small angles show some optical features, like the emission lines
 typical of FSRQs. In this category we find also ``masquerading" radio galaxies,
 that is sources, which are ``bona fide" blazars but have their non-thermal
 emission in the optical/UV part of the spectrum swamped by the host galaxy. Finally, misdirected jets
 characterize the true radio galaxy population.

 In short, starting from truly different objects like LERGs and HERGs, the combination of viewing angles and strong/weak optical/UV light dilution from the jet,
 determines the commonly adopted classification as FSRQs, BL Lacs, or radio galaxies. This purely observational approach in a number of cases assigns 
 the same class to intrinsically different objects.
 
Throughout this paper we use a $\Lambda$CDM cosmology with $H_0 = 70$ km
s$^{-1}$ Mpc$^{-1}$, $\Omega_m = 0.27$ and $\Omega_\Lambda = 0.73$
\citep{kom11}.

\section{Simulations}\label{ingredients}

Our goal is to estimate the properties of a \gr\ flux-limited blazar
sample, building upon the simulations presented in \paperone. We start by
simulating blazars in the radio band applying the same ingredients used in
\paperone\ and then estimate their \gr\ properties. To do so we must take
into account a series of constraints that are directly determined from
\fermi-LAT observations of well-defined radio selected samples of blazars.
Namely:

\begin{figure}
\includegraphics[height=7.2cm,angle=-90]{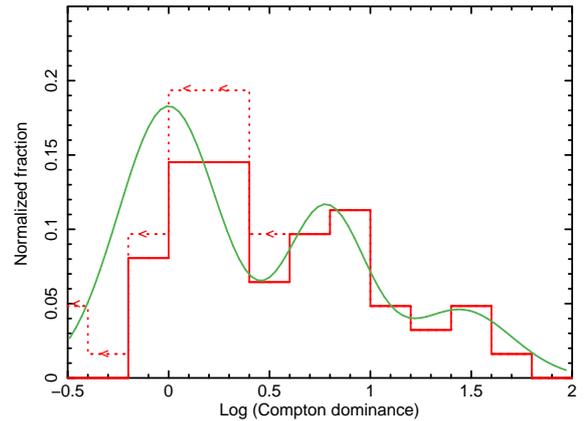}
\vspace{0.8cm}
\caption{The distribution of the logarithm of Compton Dominance in the
  sample of radio selected blazars of \citet{GiommiPlanck} with
  \nup$~<10^{14}$ Hz. The red solid histogram shows the blazars 
   detected by \fermi-LAT, while the dotted histogram represents 
  upper limits for \gr\ undetected blazars. The green line is
  the function that represents the entire population of radio selected
  blazars used for our simulations.}
\label{fig:cddistr}
\end{figure}

\begin{enumerate}

\item the redshift distributions of radio selected and \gr\, selected
  blazars are very similar \citep[see Fig. 13 of][]{fermi2lac}; moreover
  the redshift distributions of \fermi-detected and undetected radio
  selected blazars are also not significantly different \citep{lister2011};
 
\item a significant fraction ($\sim 30 - 40\%$) of radio selected blazars
  (mostly FSRQs) are not detected even in the deepest \fermi-LAT
  observations \citep{lister2011,GiommiPlanck};

\item the ratio between inverse Compton and synchrotron peak luminosities,
  usually known as Compton Dominance (CD), is crucial to predict the \gr\
  intensity of a blazar from its radio flux. \cite{GiommiPlanck} showed
  that the CD distribution of radio selected blazars is very different
  below and above \nup\ $= 10^{14}$ Hz, where \nup\ is the peak frequency
  of the synchrotron power. Fig. \ref{fig:cddistr} (adapted from Fig. 22 of
  \cite{GiommiPlanck}) shows this distribution for blazars with
  \nup$~<10^{14}$ Hz. The red solid histogram refers to all radio selected
  blazars that were detected by \fermi-LAT. The dotted histogram refers
  instead to upper limits derived from the Spectral Energy Distribution
  (SED) of those blazars that were not detected even after 27 months of
  \fermi-LAT observations \citep[see][for details]{GiommiPlanck}.  These
  undetected \gr\ sources represent a significant fraction of the sample
  and certainly cannot be ignored. The green line is our educated guess of
  the underlying intrinsic distribution. The CDs of blazars with
  \nup$~>10^{14}$ Hz (which are all detected in the \gr\ band) are instead
  clustered around $\sim 1$ \citep[see Fig. 22 of][]{GiommiPlanck}. We
  assume here that their distribution is represented by a gaussian with
 $\langle CD \rangle$ = 0.9 and $\sigma =$ 0.4. 

\item the sensitivity of \fermi-LAT is a strong function of the \gr\
  spectral index $\Gamma$ \citep[see, e.g., Fig. 14 of][]{fermi2lac}. We
  therefore need to to estimate the latter parameter as well. Since we know
  that for blazars this correlates strongly with \nup~(e.g., Fig. 29 of
  \cite{abdosed} and Fig. 17 of \cite{fermi2lac}), we use that dependence
  for our purposes.  By fitting the data of Fig. 17 of \cite{fermi2lac} we
  derive $\Gamma = -0.22 \times \log($\nup$) +5.25$, with a dispersion
  $\sigma = 0.15$.

\end{enumerate}

\subsection{The dependence of \gr\ flux on radio flux density}\label{gammaradio}

The dependence of \gr\ flux on radio flux density has been discussed for
quite some time \citep[e.g.,][and references therein]{ackermann2011}. In
this paper we will base all our calculations on the observational
constraints listed above. Based on point (i) above we therefore assume that
the \gr\ to radio flux density ratio, $f_\gamma/f_{\rm r}$, is independent
of redshift, and, in line with the minimal assumption approach of Paper I
(and consistently with observational data), we assume that this ratio is
also independent of luminosity; hence $f_\gamma/f_{\rm r}=F(CD,~$\nup$)$.

Since the CD distribution depends on \nup, for \nup$~< 10^{14}$ Hz we use a
linear dependence on CD with parameters derived from the sample of radio
selected blazars: $\log f_{\gamma}/f_{5GHz} = 0.5 \times \log(CD) -8.0$
(where $f_{\gamma}$ is in units ph~cm$^{-2}$~s$^{-1}$ and $f_{5GHz}$ is in
Jansky).  For higher \nup~values, where inverse Compton scattering occurs
in the Klein-Nishina regime and all sources have CD $\approx 1$, we use
instead a correlation between the \gr\ to radio flux ratio and \nup, that
is $\log f_{\gamma}/f_{5GHz} = 0.13 \times \log($\nup$) -9.8$.

\subsection{Simulation steps}

Our Monte Carlo simulations consist of the sequential execution of the
following steps:

\begin{enumerate}

\item all those described in detail in Paper I to simulate the radio
  through X-ray SED of blazars. In short, these include (readers familiar
  with Paper I can skip directly to point (ii)): a) drawing values of the
  radio luminosity and redshift based on our assumed luminosity function
  (LF) and evolution (see below); b) drawing a value of the Lorentz factor
  of the electrons radiating at the peak of the synchrotron power and,
  based on that, calculate \nup, assuming a simple synchrotron self-Compton
  model (with B = 0.15 Gauss and a Doppler factor $\delta$ randomly drawn
  from a gaussian distribution with $\langle \delta \rangle =15$ and
  $\sigma$= 2); c) calculating the observed radio flux density and the
  non-thermal emission in the optical and X-ray bands; d) adding an accretion
  component (only for beamed FR II sources), re-scaling an SDSS quasar
  template to a value, which depends on radio power; e) drawing a value of the
  EW of Ly$\alpha$, C~IV, C~III, Mg~II, H$\beta$, H$\alpha$
  starting from the EW distribution of the broad lines of SDSS radio quiet
  QSOs; f) adding the optical light of the host galaxy assuming a standard
  giant elliptical as observed in blazars; g) calculating the total optical
  light and the observed EW of all the broad lines by taking
  into account the dilution due to the non-thermal and host galaxy optical
  light. In order to keep our simulations as simple as possible, we used no joint probability distributions during the  process. 
  In fact, as already discussed in Paper I, some of the observed correlation between different blazar quantities could be well reproduced
  as selection effects in a simple scenario with no intrinsic correlations;
  
\item as regards the LF, since the radio powers of \gr-selected sources
  reach lower values than the radio-selected ones, we extrapolated the
  radio LF of Paper I, that is $\Phi(P) \propto P^{-3}$ (in units of
  Gpc$^{-3}$ P$^{-1}$) by a factor $\sim 4$ down to $5.0 \times 10^{23}$
  W/Hz assuming the same slope. As for the evolution, we use the same model
  of the type $P(z) = (1+z)^{k+\beta z}$, which allows for a maximum in the
  luminosity evolution followed by a decline, with $k = 7.0$ and $\beta =
  -1.5$ (which implies a peak at $z \sim 1.78$). This is only slightly
  different (within $1 \sigma$) from the $k$ value used in Paper I ($k =
  7.3$, while $\beta$ is the same) but it gives a better agreement with the
  \gr~observables without affecting the results of Paper I, as discussed in
  Sect. 5.4 of that paper.

  As detailed in Paper I, we take into account the well-known observational result that low-power radio
sources display a much weaker cosmological evolution than high-power ones by having a fraction of 
non-evolving sources equal to 1 for $P_r \le 5 \times 10^{24}$ W Hz$^{-1}$ and decreasing monotonically 
with power to 0 for $P_r \ge 5 \times 10^{27}$ W Hz$^{-1}$; 

\item for sources with \nup ~$< 10^{14}$ Hz draw a value of the CD from the
  distribution shown as a green line in Fig. \ref{fig:cddistr} and then use the 
  $f_\gamma/f_{\rm r}$ -- CD correlation, while for those with 
  \nup~$> 10^{14}$ Hz use the $f_\gamma/f_{\rm r}$ -- \nup~correlation
  (Sect. \ref{gammaradio}). Get $f_{\gamma}$ by multiplying the resulting 
  $f_{\gamma}$/f$_{5GHz}$ by the radio flux density calculated
  above;
  
\item obtain a value of the \gr\ spectral index from the correlation with
\nup~described in Sect. \ref{ingredients}; 

\item apply the \fermi~2LAC sensitivity limits, which are a strong function
  of the \gr\, spectral index as shown in Fig. 14 of \cite{fermi2lac}. The
  same plot for our simulation is shown in Fig. \ref{fig:spindvsflux}. 

\end{enumerate}

\begin{figure}
\includegraphics[height=8.2cm,angle=-90]{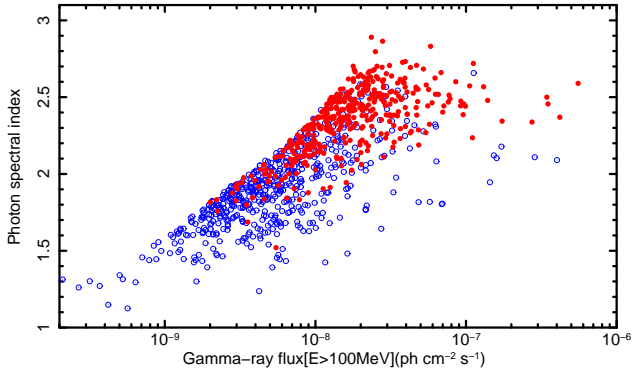}
\caption{The spectral index of blazars in our simulated survey of 1,000 \gr\ sources
 is plotted against \gr\ flux. FSRQs are plotted as filled red circles, BL Lacs as open blue circles.
  The flux cut applied reproduces the sensitivity limits of
  \fermi-LAT in the 2LAC catalogue \citep[see Fig. 14 of][]{fermi2lac}. }
\label{fig:spindvsflux}
\end{figure} 

As done in Paper I, we classify a source as an FSRQ if the rest-frame EW of at
least one of the broad lines that enter the optical band in the observer
frame (which we assume to cover the $3,800 - 8,000$ \AA~range) is $>
5$~\AA. Otherwise, the object is classified as a BL Lac, unless the host
galaxy dominates the optical light causing the Ca H\&K break to be larger
than 0.4, in which case the source is classified as a radio galaxy. A BL
Lac whose maximum EW is $< 2$ \AA, or for which the non-thermal light is at
least a factor 10 larger then that of the host galaxy, is deemed to have a
redshift which cannot be typically measured.

\section{Results: comparing simulations with \fermibf~\gr\ data}\label{results}

We now make a detailed comparison between the results of our simulations
and the observational data from the 2LAC ``clean sample" \citep{fermi2lac}
in terms of fractions of BL Lacs and FSRQs, and redshift, \nup, and \gr\
spectral index means and distributions. To include the latest measurements we have updated
the optical classifications and redshift estimates of the  ``clean sample" using the recently published spectroscopy results of \cite{shaw13}. As stressed in Paper I, the aim of
our simulations is {\it not} to reproduce the fine observational details.
Our approach is instead to reach robust conclusions keeping the number of
assumptions to a minimum.
Our simulations include 1,000 \gr\ detected blazars, a number similar to that observed in the 2LAC catalog.
To ensure statistical stability of the results the simulation was repeated ten times and the results 
were averaged.  
\begin{table}
\caption{Results from a simulation of a \gr~survey}
\begin{tabular}{llccc}
 Source type & Number of & &  \\
   & sources$^{*}$ &  $\langle z \rangle$$^{*}$ &  $\langle \log$ \nup $\rangle$$^{*}$&   $\langle \Gamma \rangle$$^{*}$\\
\hline
FSRQs      &    ~349.4        &  1.23 & 13.15 & 2.34 \\
BL Lacs     &     ~549.6 (277.3)$^{**}$   &  0.58 & 14.99 & 1.89 \\
Radio galaxies    &     ~101.0     &  0.06 & 12.98 & 2.39  \\
\hline
Total     &    1,000.0       &  0.82 & 14.15 &  2.10 \\
\hline
\multicolumn{5}{l}{\footnotesize $^{*}$Average of 10 runs each simulating 1,000 sources}\\
\multicolumn{5}{l}{\footnotesize $^{**}$BL Lacs with measurable redshift}
\end{tabular}
\label{tab:gammasim}
\end{table}

\begin{table}
\caption{2LAC data}
\begin{tabular}{llccc}
 Source type & Number of & &  \\
   & sources &  $\langle z \rangle$ &  $\langle \log$ \nup $\rangle$  &   $\langle \Gamma \rangle$ \\
\hline
FSRQs      &     309   &  1.18 & 13.23 (72\%$^{*}$) & 2.36 \\
BL Lacs     &    418 (168)$^{**}$  &  0.44 & 15.05 (76\%$^{*}$) & 1.98  \\
Unclass. blazars & 128  (14)$^{**}$ &  0.32 & 15.14 (52\%$^{*}$) & 2.07  \\
\hline
\multicolumn{5}{l}{\footnotesize $^{*}$Percentage of sources with measured \nup}\\
\multicolumn{5}{l}{\footnotesize $^{**}$Sources with measured redshift}
\end{tabular}
\label{tab:gamma2lac}
\end{table}

Table \ref{tab:gammasim} summarizes our main results by giving the number
of sources per class and their mean redshift, \nup, and \gr\ spectral
index. The number in parenthesis refers to the BL Lacs with measurable
redshift, to which the mean redshift pertain. Table \ref{tab:gamma2lac}
gives instead the observed values for the 2LAC sample, where the numbers in
parenthesis in the \nup~column give the fraction of sources for which this
parameter could be estimated.  

About 61\% of our blazars are classified as BL Lacs. This agrees well with
the observed value, which ranges between $56\pm4\%$ and
$64\pm4 \%$ \citep[where the $1\sigma$ errors are based on binomial
statistics:][]{geh86}. The first number is derived by assuming that the
unclassified blazars in \cite{fermi2lac} are all FSRQs while the second one
presumes that they are all BL Lacs. Given that their mean
\nup~and \gr\ spectral index are closer to those of BL Lacs, the
second value appears more realistic.

\begin{figure}
\includegraphics[height=7.8cm,angle=-90]{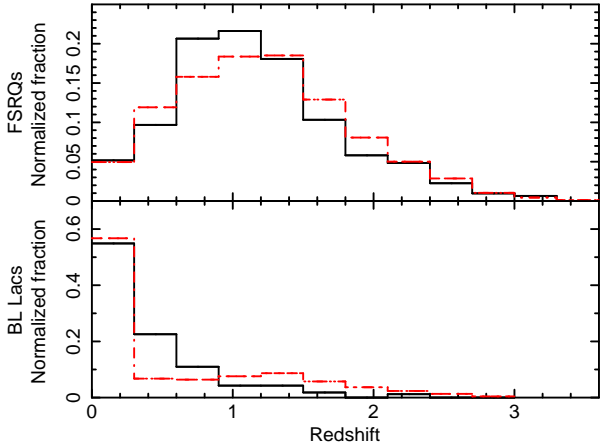}
\caption{The redshift distribution of the blazars in the 2LAC catalogue 
  \citep{fermi2lac,shaw13} (black solid histogram) and that of our simulations
  (red dot-dashed histogram) for FSRQs (top panel) and BL Lacs (bottom
  panel).}
\label{fig:zdists}
\end{figure}

The 2LAC mean value is also in good agreement with the mean redshift for
our simulated FSRQs (1.18 vs. 1.23), while for BL Lacs it is somewhat 
smaller (0.44 vs.  0.58). Fig. \ref{fig:zdists} shows the overall 
agreement between our simulated redshift distributions (where we have only
included sources with a measurable redshift) and the observed ones. 

$70\%$ of the BL Lacs ($44\%$ of those with redshift) have a standard
accretion disk and are therefore FSRQs, which are classified as BL Lacs
only because their observable emission lines are swamped by the non-thermal
continuum, as was the case for many radio-selected blazars in Paper I. 

About 50\% of our BL Lacs have a redshift determination, which is close to
the 2LAC value of $ 40\pm4 \%$ \citep{shaw13}. As stated in \cite{fermi2lac}, many of the
unclassified blazars do not have an optical spectrum so we cannot include them
in the evaluation of the fraction of BL Lacs for which a redshift could not
be obtained because they were totally featureless. According to our
simulations, the ``real" redshift of featureless BL Lacs ranges typically
between 0.5 and 2.5, which is much larger than that of those with redshift.
Basically all of these sources possess an accretion disk and therefore
are FSRQs.

A small fraction ($\sim 10 \%$) of the simulated blazars are classified as
radio galaxies, again as was the case in Paper I. These are bona-fide
blazars misclassified because their non-thermal radiation is not strong
enough to dilute the host galaxy component. Their relevance is discussed
in Sect. \ref{fakeRGs}.

\begin{figure}
\includegraphics[height=8.0cm,angle=-90]{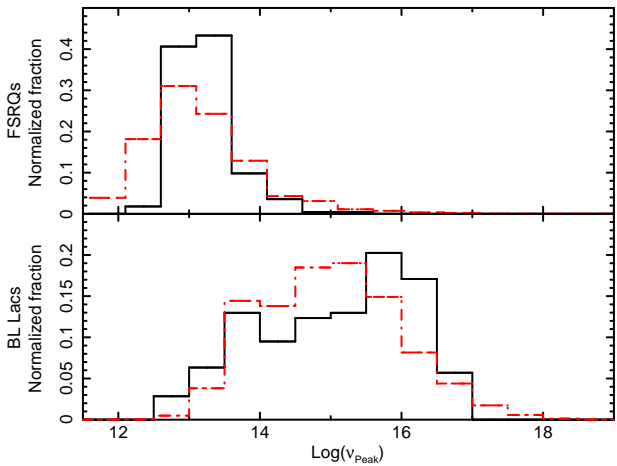}
\caption{The distribution of synchrotron peak energies for blazars in the
  2LAC catalogue (black solid histogram) and that of our simulations (red
  dot-dashed histogram) for FSRQs (top panel) and BL Lacs (bottom panel).}
\label{fig:nupdist}
\end{figure}

The predictions for \nup~agree well with observations, as shown in Tab. 
\ref{tab:gammasim} and \ref{tab:gamma2lac}. The 2LAC mean value
is $\langle \log$ \nup $\rangle = 15.07$ if one includes the
unclassified blazars with the BL Lacs. Fig. \ref{fig:nupdist} compares the
distributions of \nup~for sources classified as FSRQs and BL Lacs in our
simulation with those of the 2LAC blazars. The agreement is clearly quite
good and reproduces the well known fact that \gr-selected BL Lacs tend to
have \nup~values significantly higher than FSRQs
\citep{abdosed,GiommiPlanck}. One caveat here is that only about 3/4 of
2LAC blazars have \nup~values, which moreover were not estimated from a fit
to the SED but by using the radio-optical and X-ray-optical broadband
indices.

The \nup~values for our simulated BL Lacs without redshift peak at $\sim
10^{15}$ Hz, as for the 2LAC sources. Only $\sim 10\%$ of
simulated BL Lacs are Low Synchrotron Peaked (LSP: \nup~ $< 10^{14}$~Hz),
with $\sim 33\%$ being Intermediate Synchrotron Peaked (ISP: $10^{14}$~Hz
$<$ \nup~$< 10^{15}$~Hz) and $\sim 57\%$ High Synchrotron Peaked (HSP:
\nup~$> 10^{15}$~Hz). The corresponding 2LAC values are $\sim 20\%$, $\sim
25\%$, and $\sim 55\%$, which are not too different, considering also the
caveat mentioned above. Very similar fractions ($\sim 23\%$, $\sim 20\%$,
and $\sim 57\%$) are obtained by including with the BL Lacs the
unclassified blazars.

\begin{figure}
\includegraphics[height=8.0cm,angle=-90]{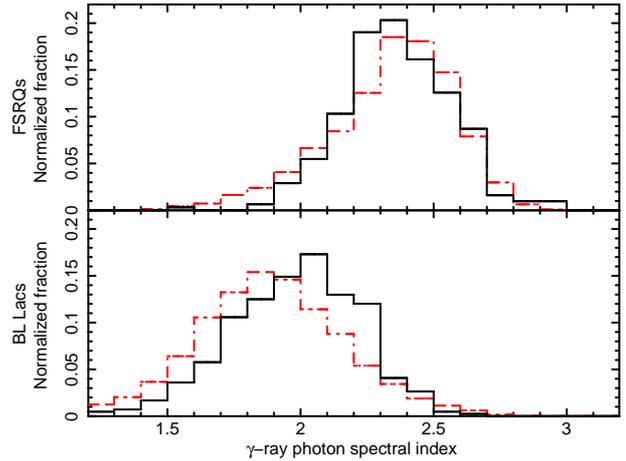}
\caption{The \gr\ spectral (photon) index distribution of FSRQs (top panel)
  and BL Lacs (bottom panel). Black solid and red dot-dashed histograms
  represent 2LAC and simulated blazars respectively.}
\label{fig:spindistr}
\end{figure}

Finally, the \gr\ spectral photon indexes are also consistent with
observations, with a 2LAC mean value of $\langle \Gamma \rangle = 2.00$ if
one includes the unclassified blazars with the BL Lacs. Fig.
\ref{fig:spindistr} compares the distributions for our simulated FSRQs and
BL Lacs with those of the 2LAC blazars. Despite the slight shift in the
peak of the distribution for BL Lacs, the agreement is again quite good and
reproduces the well known fact that \gr-selected BL Lacs tend to have \gr\
spectral indexes flatter than those of FSRQs \citep[e.g.][and references
therein]{fermi2lac}.

\subsection{\gr\ predictions for radio selected blazars}\label{radiosurvey}

To further check our scenario and verify that we are consistent with the
observational constraints listed in Sect. \ref{ingredients}, we have
simulated a sample of radio selected blazars with a flux density limit of
1 Jy. We have then compared our \gr\ predictions with the \fermi-LAT results
for the sample of radio bright blazars reported in \citet{GiommiPlanck}.
We find that about 71\% and 91\% of simulated radio selected FSRQs and BL
Lacs respectively are above the \fermi\ 2LAC sensitivity limit. These
percentages compare very well with the observed \fermi-LAT detection rate
for FSRQs (71$^{+16}_{-13}$\%) and BL Lacs (100$^{+0}_{-33}$\%) in the
radio sample.

\citet{fermi2lac} showed that the redshift distribution of \gr\ selected
blazars is very similar to that of bright radio selected blazars (see their
Fig. 13 for a comparison of the 2LAC and the WMAP blazar samples).  The
redshift distributions of our simulated radio and \gr\ selected blazars are
also very similar with average values close to the observed values $\langle
z \rangle _{radio}$ = 1.2 and $\langle z \rangle _{\gamma-ray}$ = 1.2 for
FSRQs, and $\langle z \rangle_{radio }$ = 0.7 and $\langle z \rangle
_{\gamma-ray}$ = 0.6 for BL Lacs with redshift determination. We make
one more comparison between simulations and observations in
Sect. \ref{gamma_slope_plane}. 

Given the similarities in the redshift distributions and considering that the 
\gr\ to radio flux density ratio is independent of redshift, we expect the 
cosmological evolutionary properties of radio and \gr\ selected blazars to be 
the same. 


\subsection{Assessing the stability of our results}

As we did in Paper I we assessed the dependence of our results on the
adopted LF and evolution by making two checks: first, we
varied their input values by $1 \sigma$; second, we used as an alternative
LF the sum of the BL Lac and FSRQ LFs based on the beaming model of
\cite{UP95} (converted to $H_0 = 70$ km s$^{-1}$ Mpc$^{-1}$) assuming, for
consistency with the way they were derived, a pure luminosity evolution of
the type $P(z) = P(0) {\rm exp} [T(z)/\tau]$, where $T(z)$ is the look-back
time, and $\tau = 0.33$ \citep[consistent with the evolution of DXRBS FSRQs
and BL Lacs combined, based on the samples in][]{pad07}.
We  also tested the sensitivity of our simulations to different assumptions  
on the Doppler factors using distributions with mean values that ranged between
5 and 30 depending on the radio luminosity of the simulated blazar. 

In all cases our main results, that is the prevalence of BL Lacs in \gr\ selected
samples and the higher redshifts, lower \nup, and larger \gr\ spectral
indexes of FSRQs were confirmed, which shows that they are independent of
the details of the LF, evolution, and assumed Doppler factor.



\section{Discussion}
 
\subsection{\fermibf~BL Lacs without redshift}

Despite extensive spectroscopic observations $56\pm5\%$ of the BL
Lacs in the 2LAC do not have a redshift determination due to the lack of
features in their optical spectrum \citep{fermi2lac}. This high fraction of
redshift-less blazars is matched very well by our simulations, which indeed
predict a value $\sim  50\%$. Most of these sources are of the HSP type,
$\sim 1/3$ are ISP, while only $\sim 1/10$ are LSP. These
fractions are also in agreement with what is observed in the 2LAC
sample.

Our simulations imply that these sources have much
larger redshifts than those of BL Lacs with redshift information: while in the latter case 
 $\sim 62\%$ of objects have redshift $\lsim 0.5$ , in the former case redshift values range mostly between 0.5 and 2.5, with peak probability between 0.9 and 1.2. 
Therefore, one cannot assume for them a value typical of BL Lacs \citep[as done 
by][]{fermi2lac} but larger values ($z \approx1$) need to be used. Moreover, most
of these sources are predicted to be quasars with their emission lines
heavily diluted by the non-thermal continuum. These should therefore be
included with the FSRQs when studying number counts, cosmological
evolution, and LFs, since their exclusion will bias the results.

\begin{figure}
\includegraphics[height=8.cm,angle=-90]{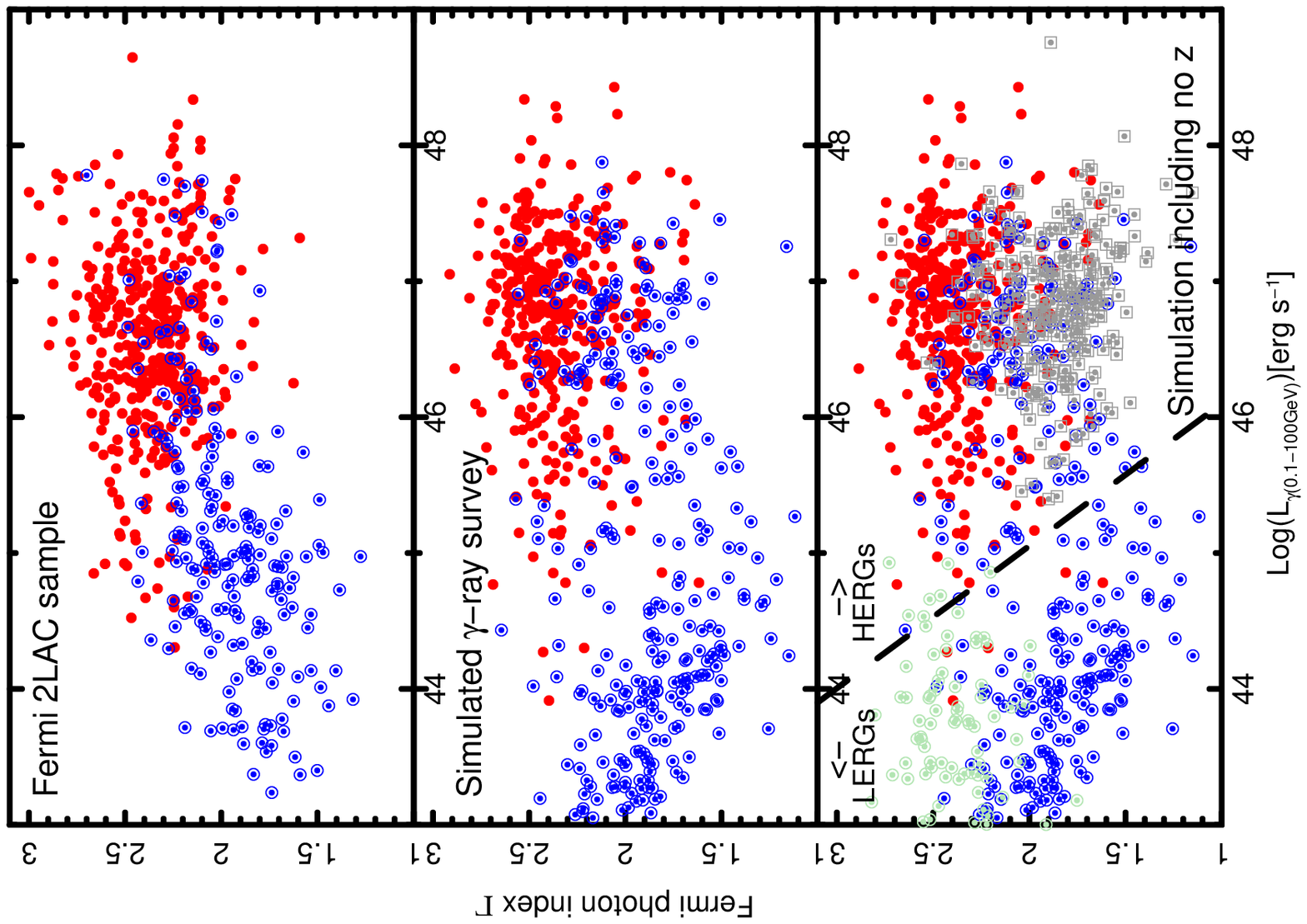}
\caption{The \gr\ spectral index vs. \gr\ luminosity plot for \gr\ flux
  limited samples; FSRQs: filled red circles, BL Lacs: open blue circles.
  Top panel: the case of the 2LAC catalogue \citep[][]{fermi2lac} ; middle
  panel: simulated \gr\ flux limited survey where objects with no redshift
  information are not plotted; bottom panel: simulated \gr\ survey where
  the blazars with no redshift determination appear as gray open squares
  and sources classified as radio galaxies appear as light green
  symbols. 
  The dashed line gives the approximate boundary between LERG and HERG blazars in the simulation.
 }
\label{fig:glfvsesp}
\end{figure}

\begin{figure}
\includegraphics[height=8.cm,angle=-90]{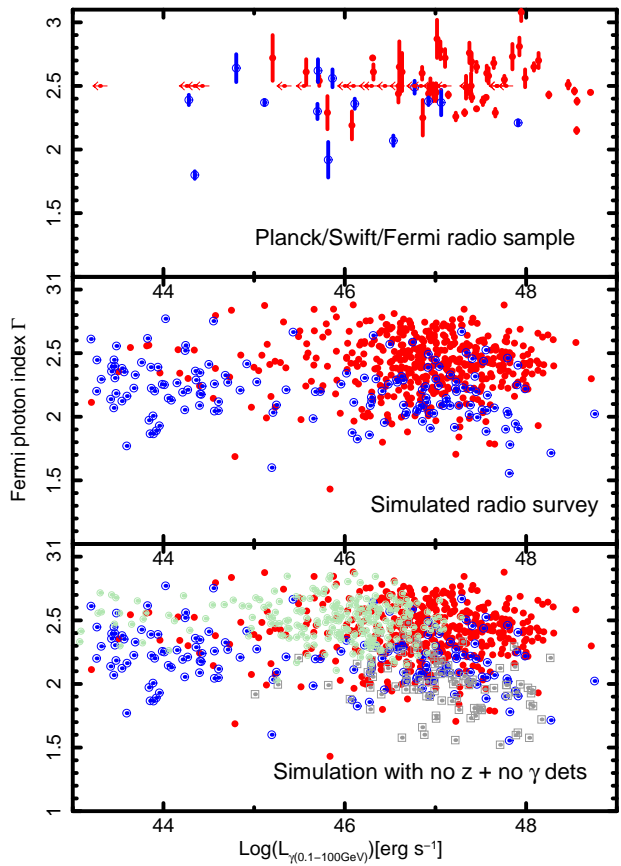}
\caption{The \gr\ spectral index vs. \gr\ luminosity plot for radio flux
  limited samples; FSRQs: filled red circles, BL Lacs: open blue circles.
  Top panel: the case of the Planck Swift Fermi radio bright sample
  \citep[data taken from][]{RadioPlanck}; middle panel: simulated radio
  flux limited survey where objects with \gr\ flux below the 2LAC catalogue
  limit and those with no redshift information are not plotted; bottom
  panel: simulated radio survey where the blazars with \gr\ flux below the
  2LAC limits appear as light green points and sources with no redshift
  determination appear as gray open squares. 
 }
\label{fig:rlfvsesp}
\end{figure}

We note that the existence of such sources has been recently confirmed at
relatively high redshift. \cite{pad12} have studied the SEDs of eleven 2LAC
blazars without spectroscopic redshift but for which a photometric redshift
$> 1.2$ was derived by fitting SED templates to their UV-to-near-IR
multi-band photometry obtained quasi-simultaneously with {\it Swift}/UVOT
and GROND by \cite{rau12}. Four blazars turned out to be of a type never
seen before but with properties expected by us for some of the BL Lacs
without spectroscopic redshift: large \nup~($\sim 5 \times 10^{15}$ Hz) and
high-power ($L_{\rm peak} \approx 6 \times 10^{46}$ erg s$^{-1}$). Four
more were ISP and two were LSP, all six with $L_{\rm peak} \ga 2 \times
10^{46}$ erg s$^{-1}$, which is well into the FSRQ regime. It then looks
like these sources are very plausible FSRQs in disguise.

\subsection{``Masquerading" BL Lacs}

How can one, in practice and in general, spot the ``masquerading" BL Lacs? There
are a few options:

\begin{enumerate}
\item if a photometric redshift estimate is available, as in the case of
  the sources studied by \cite{rau12}, one can derive the radio, or
  synchrotron peak, power and see if they are FSRQ-like;

\item if infrared spectroscopy is available then there is a good chance
  that the source could turn out to be an FSRQ. Our simulations of the 2LAC,
  in fact, imply that $\approx 50\%$ of the BL Lacs without redshift should
  show a strong (EW $> 5$ \AA) H$\alpha$ line, which for $z \sim 1$ would
  fall, for example, at $\sim 1.3$ $\mu$m. For a further $\approx 25\%$ one
  should find $2 < \rm{EW} < 5$ \AA~and therefore it should be possible to
  assign a redshift to these sources as well;

\item finally, as mentioned in Sect. \ref{results}, basically all sources
  without redshift in the 2LAC catalogue are predicted by our simulations
  to have an accretion disk and therefore to be intrinsically quasars. So
  for this sample the mere fact of having a featureless spectrum should be
  enough for a source to be considered, perhaps ironically, an FSRQ with high
  probability. We stress that this result is sample- and flux density 
  limit-dependent.
  \end{enumerate}
  
  In summary, we notice that in our scenario the historical 5
  \AA~separation between BL Lacs and FSRQs has no physical meaning and that
  the detection of {\it any} line associated with the broad line region of
  {\it any} strength should be sufficient to grant a source FSRQ status.

\subsection{The \gr\ - spectral slope luminosity plane}\label{gamma_slope_plane}

\cite{ghis11,ghis12} noted a strong correlation between the 100MeV -- 10GeV
luminosity and the \gr\ spectral slope in the 2LAC sample,
arguing that this relationship reflects a physical distinction between
FSRQs and BL Lacs. In this interpretation the presence (or absence) of
strong emission lines is associated to a regime of high (low) radiating
efficiency and, consequently, larger (or smaller) \gr\ luminosity.

We tested if such a correlation could be induced by the \fermi-LAT
sensitivity dependence on spectral slope by comparing the 2LAC data with our
simulated \gr\ survey blazars. Fig. \ref{fig:glfvsesp} plots the 
\gr\ slope vs. 100MeV -- 10GeV luminosity for blazars with a measurable
redshift in the 2LAC clean sample (top panel), and in our simulated survey
(central panel). The trend of high-luminosity/steep \gr\ spectrum
associated to FSRQs (filled red points) and low luminosity/flat \gr\
spectrum associated to BL Lacs (open blue points) is clearly present also in
the simulation, with BL Lacs mostly concentrated in the flat-spectrum/low
luminosity corner 
and FSRQs populating almost exclusively the steep
spectrum/high luminosity part of the plot. If we consider also BL Lacs with
no redshift determination and objects (mis)classified as ``radio galaxies"
(gray squares and light green open circles respectively in the bottom panel
of Fig. \ref{fig:glfvsesp}) we see that the \gr\ - spectral slope
luminosity plane is filled in all its quadrants, although the different
densities imply that a strong correlation is still present. 
We note that below $\sim 10^{44}$ erg s$^{-1}$ we predict 
the existence of BL Lacs with photon indexes somewhat steeper than those observed
in the 2LAC sample. This is likely due to our
simple power law extrapolation of the radio LF  and to the misclassification  
of some sources close to the BL Lac/``radio galaxy" border. The LF extrapolation is
also behind the somewhat higher fraction of simulated sources
with $\lsim 10^{45}$ erg s$^{-1}$ as compared to the
observed one.

We then conclude that the trend reported by \cite{ghis11,ghis12} comes out
naturally from our simulations and reflects the complex \fermi-LAT
sensitivity limits.
However, we also note that the separation between ``real" (that is, LERG)
BL Lacs and FSRQs, that in our simulation fall to the left and to the right of the dashed line of Fig. \ref{fig:glfvsesp},
is in agreement with the deductions of \cite{ghis12} for sources with known redshift. In our scenario 
BL Lacs with no redshift  are almost all expected to be to the right of the dashed line, and therefore to be HERGs and diluted FSRQs.

To better understand the \gr\ spectral slope - luminosity plane we
simulated the \gr\ properties of a sample of blazars with a {\it radio flux
  density} limit equal to those of the sample considered by
\cite{RadioPlanck} and compared our expectations with the \fermi-LAT
measurement reported by \cite{GiommiPlanck}.

Fig. \ref{fig:rlfvsesp} is the same as Fig. \ref{fig:glfvsesp} for the case
of a radio flux density limited sample (see Sect. \ref{radiosurvey}). The top
panel plots the \gr\ spectral slope versus luminosity for the sample
described in \cite{RadioPlanck} and \cite{GiommiPlanck}. The left pointing
arrows represent blazars not detected even in the 27 month \fermi-LAT data
to which a spectral slope of 2.5 was assigned. BL Lacs with no
redshift determination are not plotted. The results of our simulations are
shown in the middle and lower panels where we plot respectively FSRQs and BL
Lacs expected to be detected by \fermi-LAT in a 27 month integration as
red and blue points, while BL Lacs with no redshift determination are
shown as gray open symbols, and blazars undetected by \fermi-LAT appear as
light green circles.

As apparent from Fig. \ref{fig:rlfvsesp} in this radio selected sample of
blazars the correlation between \gr\ spectral slope and luminosity noted by
\cite{ghis11,ghis12} is not present, in good agreement with our
simulations. This confirms that it is crucial to fully understand the
biases of the various surveys before one can draw any conclusion on the
underlying physics.

\subsection{The blazar \gr\ - radio flux density correlation}

Correlations between the blazar \gr\ and radio flux densities have been
discussed by several authors in the past few years without reaching
definite conclusions. This was due to limitations connected to the
availability of large and unbiased samples, lack of simultaneity, and
presence of a significant number of non-detections. Recently, a consensus
seems to have been reached on the existence of a correlation between the
two bands, albeit with a large scatter \citep[e.g.,][and references
therein]{ackermann2011,GiommiPlanck,tavares12}.

\begin{figure}
\includegraphics[height=7.8cm,angle=-90]{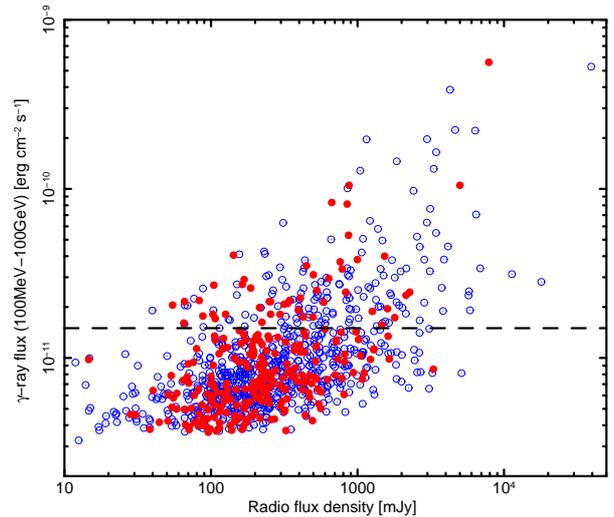}
\caption{The \gr\ flux vs. radio flux density in our 2LAC simulated survey 
(reduced to 1,000 sources for better visibility). Filled red circles represent FSRQs, open blue circles represent
  BL Lacs. The dashed line corresponds to the \gr\ flux limit of a
  survey a factor $\approx $ 5 shallower than 2LAC.}
\label{fig:gvsr}
\end{figure}

Fig. \ref{fig:gvsr} shows the \gr\ flux vs radio flux density for our
simulated \gr\ survey. This is remarkably similar to the equivalent plot
based on \fermi-LAT data \citep[Fig. 3 of][]{ackermann2011}.

Despite the large spread in $f_\gamma/f_{\rm r}$, induced by the wide
dynamical range of the observed CD distribution (which can result in \gr\
fluxes up to a factor 100 or more different for the same radio flux
density), a correlation is clearly present in the 2LAC simulation 
shown in Fig. \ref{fig:gvsr}. For \gr\ surveys with much
shorter integration times (e.g. corresponding to the dashed line in Fig.
\ref{fig:gvsr}), the \gr\ dynamical range decreases significantly while the
radio one remains similar. This, combined with the much reduced statistics, 
makes it much more difficult to detect the intrinsic \gr-radio correlation.

\subsection{\fermibf-LAT unassociated high-Galactic latitude sources}\label{fakeRGs}


As mentioned in Sect. \ref{results}, the scenario we propose 
predicts the existence of ``masquerading" radio galaxies, that is blazars having
their optical non-thermal emission swamped by the host galaxy. 
These are intrinsically rare sources with high CD
values that, when detected close to the \fermi-LAT sensitivity limit, have
a radio and optical non-thermal component significantly lower than that of blazars with
smaller CD. They would then display a pure elliptical galaxy optical
spectrum and would appear as radio sources typically fainter than currently
known \fermi\ steep spectrum blazars (that are also characterized by large LAT error regions) thereby failing the association criteria applied by
\cite{fermi2lac}. 
Objects of this type could then be the counterparts of a
significant fraction of the still unassociated high-latitude \fermi\ \gr\
sources, which make up $\sim 20\%$ of the total at $|b| > 10^{\circ}$. This
is the same mechanism at work in shallow X-ray surveys where intrinsically
rare objects with high $f_{x}/f_r$, that is HSP, are preferentially
selected over the more common LSP.

\begin{figure}
\includegraphics[height=8.0cm,angle=-90]{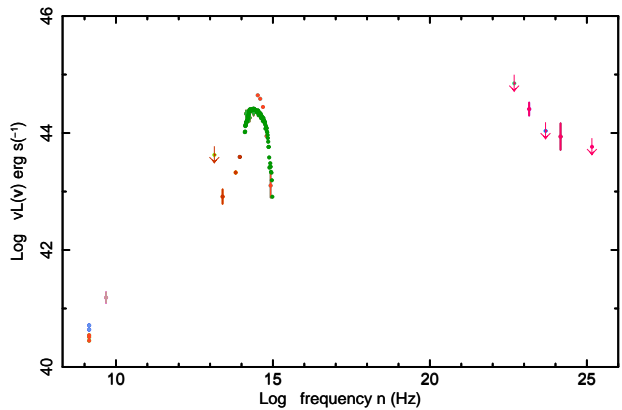}
\caption{The SED of the radio source coincident with the 17 magnitude z=
  0.238 elliptical galaxy SDSSJ173130.8+5429 within the error ellipse of
  the unassociated \fermi\ \gr\ source 2FGL J1730.8+5427. An example of a
  high CD radio galaxy as a possible counterpart of unassociated \fermi-LAT
  high galactic latitude sources. The green line represents the template of
  a giant elliptical galaxy \citep{man01}.  }
\label{fig:rgal}
\end{figure}

A systematic search for such elusive counterparts of \fermi-LAT sources is
beyond the scope of this paper and will be carried out in the future.
Fig. \ref{fig:rgal} shows, as an illustrative example, the SED of
SDSSJ173130.8+5429, which might be a source of this type within the error
ellipse of the presently unassociated \gr\ source 2FGL J1730.8+5427.






\section{Conclusions}

We have carried out Monte Carlo simulations of the radio through \gr\
emission of blazars selected in the \gr\ and radio bands within the
framework of the newly proposed {\textit {simplified blazar view} 
presented in Paper I. 
We have extended these simulations to the \gr\ domain 
using a \gr\ to radio flux density ratio that is constrained by the relationships between SED parameters discovered by \cite{abdosed} and by the distribution of
Compton Dominance derived by \cite{GiommiPlanck}. No dependence on redshift or luminosity was assumed. 

Our main results can be summarized as follows:

\begin{enumerate} 
\renewcommand{\theenumi}{(\arabic{enumi})} 
 
\item Our results match very well the blazar content and the observational
  properties of the \gr\ sources included in the 2LAC catalogue
  \citep{fermi2lac}. In particular, we reproduce all the main features of
  the 2LAC sample, namely the prevalence of BL Lacs, the larger redshifts
  and \gr\ spectral indices and lower \nup~of FSRQs as compared to BL Lacs,
  and the fraction of BL Lacs without redshift, including their split in
  terms of \nup.

\item The simulations also reproduce well the \gr\ properties of the radio
  flux density limited sample considered by \citet{RadioPlanck} and
  \citet{GiommiPlanck}.

\item Our simulations imply that the redshift of the BL Lacs with
  featureless optical spectrum is much larger than that of those with
  measured redshift, with values ranging from 0.5 to 2 and peak probability around 1. 
  These sources are also predicted to possess an
  accretion disk and therefore not to be real BL Lacs but quasars, and as
  such should be considered.

\item The tendency for BL Lacs of appearing as low \gr\ luminosity/flat spectrum sources 
and of FSRQs of being high luminosity/steep \gr\ sources noted by \citet{ghis11,ghis12}
is also present in our simulated 2LAC survey, but not in real and simulated radio flux-density limited samples.   
In both our scenario and in the interpretation of  \citet{ghis11,ghis12} low luminosity BL Lacs are 
low accretion LERG objects, while FSRQs and high luminosity steep spectrum BL Lacs are HERG sources;
however in our simulations, contrary to the assumptions of  \citet{ghis11,ghis12}, almost all the redshift-less BL Lacs are diluted FSRQs.

\item The strength and large scatter of the observed \gr\ -- radio flux density 
  correlation, which has caused some discussion in the literature, are 
  reproduced by our simulations. The latter is induced by the wide 
  dynamical range of the observed Compton Dominance distribution.  

\item About 10\% of the \gr\ sources are classified as radio-galaxies on
  the basis of their optical properties but are instead moderately beamed
  blazars with their non-thermal emission swamped by the host galaxy. These
  objects fail to pass the association criteria typically applied to \gr\
  samples and could therefore explain a significant fraction of the still
  unassociated high-latitude \fermi\ sources.
  
\end{enumerate}

\section*{Acknowledgments}

PP thanks the ASI Science Data Center (ASDC) for the hospitality and
partial financial support for his visits. We acknowledge the use of data
and software facilities from the ASDC, managed by the Italian Space Agency
(ASI). Part of this work is based on archival data and on bibliographic
information obtained from the NASA/IPAC Extragalactic Database (NED) and
from the Astrophysics Data System (ADS). We thank the anonymous referee
for useful comments.

\label{lastpage}
\end{document}